# Theoretical modeling of dendrite growth from conductive wire electro-polymerization

**Ankush Kumar[1]\*, Kamila Janzakova[1], Yannick Coffinier[1], Sébastien Pecqueur[1], Fabien Alibart[1,2]**

1  Univ. Lille, CNRS, Centrale Lille, Univ. Polytechnique Hauts-de-France, UMR 8520 - IEMN, F59000 Lille, France.
2  Laboratoire Nanotechnologies &Nanosystèmes (LN2), CNRS, Université de Sherbrooke, J1X0A5, Sherbrooke, Canada.
*ankush.kumar@iemn.fr,  ankush.science@gmail.com

## Abstract
Electropolymerization is a bottom-up materials engineering process of micro/nano-scale that utilizes electrical signals to deposit conducting dendrites morphologies by a redox reaction in the liquid phase. It resembles synaptogenesis in the brain, in which the electrical stimulation in the brain causes the formation of synapses from the cellular neural composites.  The strategy has been recently explored for neuromorphic engineering by establishing link between the electrical signals and the dendrites' shapes. Since the geometry of these structures determines their electrochemical properties, understanding the mechanisms that regulate polymer assembly under electrically programmed conditions is an important aspect. In this manuscript, we simulate this phenomenon using mesoscale simulations, taking into account the important features of spatial-temporal potential mapping based on the time-varying signal, the motion of charged particles in the liquid due to the electric field, and the attachment of particles on the electrode. The study helps in visualizing the motion of particles in different electrical conditions, which is not possible to probe experimentally.  Consistent with the experiments, the higher AC frequency of electrical activities favors linear wire-like growth, while lower frequency leads to more dense and fractal dendrites' growth, and voltage offset leads to asymmetrical growth. We find that dendrites' shape and growth process systematically depend on particle concentration and random scattering. We discover that the different dendrites' architectures are associated with different Laplace and diffusion fields, which govern the monomers' trajectory and subsequent dendrites' growth. Such unconventional engineering routes could have a variety of applications from neuromorphic engineering to bottom-up computing strategies.

## Introduction

The brain is an incredible computing system that has served as an inspiration for the construction of circuits for computational hardware and software.[1-5] Brain-inspired machine learning-based software techniques have accomplished various feats in terms of handling image, signal, and natural language-based challenges,[1-2] but they continue to suffer in terms of many orders of high energy consumption.[3] One key explanation for the computing hardware's inadequacies in comparison to the brain is its architecture.[6,7] On the one hand, the brain uses a bottom-up strategy for construction, using biochemical ingredients in liquid and electrical activity, whereas existing software and hardware, on the other hand, use a top-down approach, in which all possible connections must be created before training, and their strengths are modulated during the training process, and weak connections are only precluded.[8-9] It results in several orders of magnitude higher energy consumption of existing hardware compared to the brain due to fully connected network training, and from a hardware standpoint, it is extremely difficult to create fully connected networks with a higher number of neurons due to space constraints for dense connectivity. Thus, the biological strategy of generating just relevant connections might be a valuable option for conserving resources, space, and energy usage in computer hardware.[6, 7,10, 11] With this inspiration, the materials engineering technique of electropolymerization[12-13] has been explored to create electrical activity-driven connections.[15-20] Monomers present in the liquid phase polymerize on the electrodes based on electrical activity in these techniques.[12-13] The connection between the electrodes in this case may be managed by electrical activity between the electrodes, which is important for neuromorphic applications. We recently demonstrated experimentally that signal parameters may be utilized to consistently generate diverse interconnected morphologies.[14] Koizumi et al. showed the electropolymerization of Poly(3,4-ethylenedioxythiophene) (PEDOT) derivatives and discovered that the propagation was in the form of fiber from the ends of Au bipolar electrodes (BPEs) in the parallel direction to the external electric field.[16] By adjusting the applied voltage, duty factor, and electrode spacing, Eickenscheidt et al. showed the formation of diverse polymeric forms.[17] Unlike previous neuromorphic options, the suggested technique is more closely related to biological circumstances and is created from monomers present in the liquid phase, similar to the constituents of neurotransmitters involved in synaptogenesis.[6, 7,10, 11] With these dendritic wires, the engineering path has opened up opportunities for neuromorphic computation, and several research groups are investigating electro-polymerization as a viable technique for neuromorphic computing. In this spirit, Akayi-Kasaya et al. demonstrated the suitability of conductive polymer wires derived by bipolar electro polymerization for neuromorphic applications, where the morphology's electro polymerization development is directly related to the learning process of artificial synapses.[18] Hagiwara et al. demonstrated long-term potentiation and short-term potentiation using the varied frequency of continuous pulses.[19] Ji et al. built organic electrochemical transistors (OECTs) based on bipolar electropolymerization, and the performance of the OECTs may be modified by adjusting the electropolymerization settings.[20] Very recently, the technique has been explored for neuromorphic functions such as Hebbian learning and pattern recognition.[21] The electopolymerized dendrite structures, controlled by the parameters can also be employed in several applications needing fractal electrodes such as super capacitors[22] heat transfer[23], fractal antennas[24], fractal absorbers,[25] solar cells electrodes[26] ,and electronic wire connections[27] etc.

Since one can employ different conditions of chemical material type, experimental conditions, electrical parameters, and electrode geometry configurations, it is important to understand the phenomena, generic dependence of parameters, affecting their growth process to well tune their morphologies for desired connection types. In all of the preceding situations, the experimental investigations primarily highlight the growth and final pattern of dendritic structures under various electrical activities and experimental settings. Very recently, the potential generated in the process has been attempted to be mapped by electrochemiluminescence.[28] However, since the growth occurs from one specific composition of chemical species diluted or dispersed in the liquid phase, it is beyond the experimental and optical microscopy scope to see the motion and deposition of such ingredients forming a particular morphology and systematically predict it under various chemical compositions. There are currently no modeling methodologies available to describe the evolution of electro-polymerized structures under diverse electrical, chemical and geometrical settings. Ab-initio simulations based on atomic and molecular interactions are not capable of modelling structures with orders of magnitude larger than molecular sizes. [12,13,16] Furthermore, the time-varying signal and changing electrode make such simulations challenging to run on a commercial tool. Taking these considerations into account, we present a mesoscale simulations model that incorporates particle interaction and mobility, based on the assumption that mesoscale dendritic morphologies are yielded by charged particle aggregation of oligomers preformed and dispersed in the electrolyte prior to seeding on the electrode. The modeling technique can be useful in understanding the influence of required elements on morphology and in establishing the link between electrical signal and dendrite-morphology. The modeling provided in the paper reflects the many forms of neurons and synapses that are regulated by a few factors such as growth area, pruning time scale, and spatial distribution of 'neurotrophic particles.'[29] Apart from the neuromorphic engineering community, the investigation can be an important contribution to different domains of science interested in pattern formation. As an example, uniform to fractal like atomic growth with growth conditions of Molecular Beam Epitaxy[30], coffee ring effects[31], based on liquid evaporation conditions and investigation of different morphologies with dc- electrodeposition[32-33].

With this in perspective, we attempted to study how electrical factors may be utilized to program Conducting-Polymer Dendritic Interconnections and regulate their morphology. The modeling entails the spatiotemporal mapping of the electric field in the liquid as regulated by the electrical signal applied to the electrodes, motion of the constituent particles in the field, and electro-polymerization on both electrodes, resulting in variable morphologies under different conditions. The paper consists of three sections: (i) Proposed modeling methodology, (ii) Comparison of modeling results with experimental observations, and (iii) modeling predictions.

**Results and Discussion:**

**Proposed modeling methodology:**

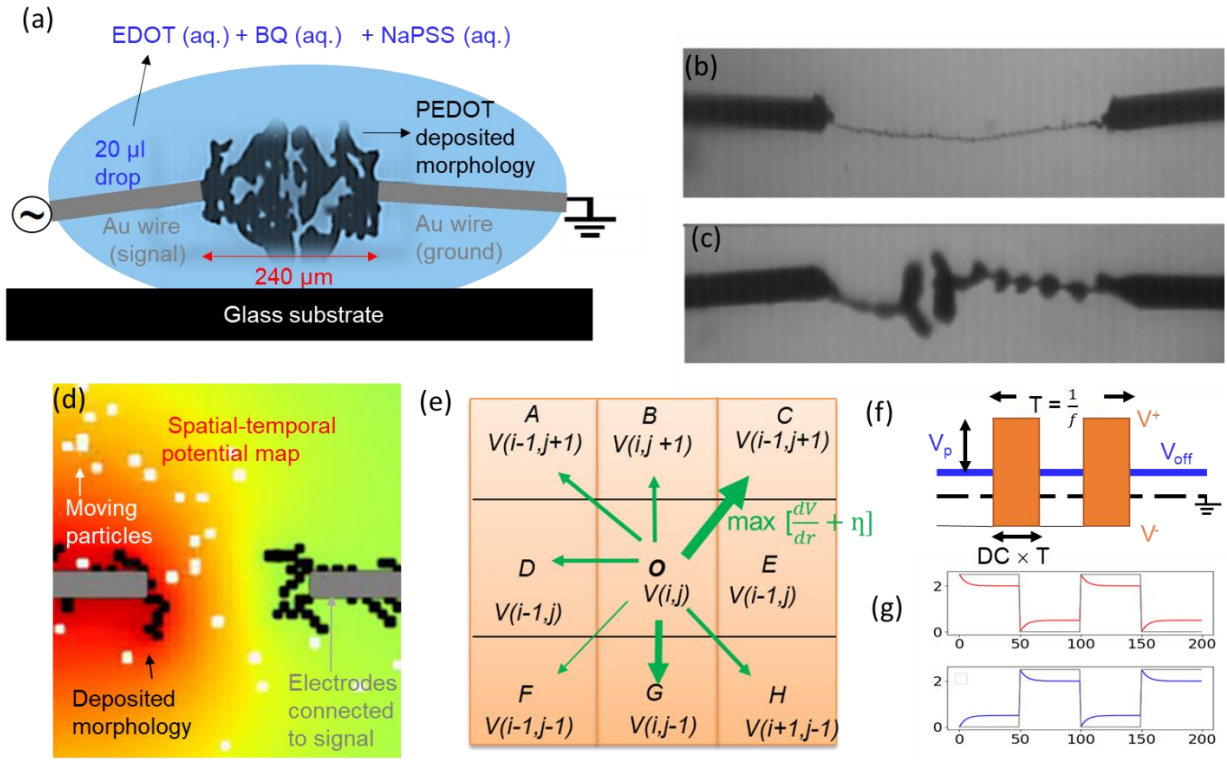

Figure 1: (a) Experimental setup for bipolar electro polymerization of PEDOT consisting of an aqueous drop containing monomers of PEDOT (EDOT), benzoquinone (BQ) as an oxidizing agent and sodium polystyrene sulfonate (NaPSS) as electrolyte and dopant for the dendritic microstructures. A specific periodic signal is applied on one of the Au wire electrodes, while the other Au wire electrode is at the ground. (b-c) morphologies of polymerized materials formed for different electrical signals.[14] (d) Simulation geometry for electro-polymerization by considering motion and attachment of particles on the electrodes. The complete box represents the liquid, gray lines represent the electrodes connected to specific signal, the white points represent the moving particles, and black points represent the polymerized particles on the electrodes. The spatiotemporal potential map is evaluated based on the applied waveform and modified electrode, simulation parameters are listed in the table. (e) The motion of a particle is driven by combined contribution of scattering and electric field motion. The particle is made to move in any direction with the probability based on effective forces. (f) Signal

**applied on the electrodes. (g) With the voltage signal (shown in gray), the voltage experienced at the dielectric surface is calculated based on capacitive-resistance time scales (shown in blue and red).**

Figure 1a represents the experimental setup of the electropolymerization technique, with a drop of aqueous electrolyte containing EDOT monomers, NaPSS electrolyte and benzoquinone oxidizing agent. Dendrites were grown by bipolar alternating current with one electrode connected to time varying signal and other connected to ground. The characteristics of time varying signal can be used to modulate the geometry of dendrites. One can achieve wire-like, fractal, engulfing nature of dendrites by playing with electrical parameters such as applied voltage, frequency, duty cycle and offset etc.[14] (see Figure 1b-c for few examples). From the experimental perspective, one can only perform the imaging of the deposited conducting polymer morphology occurring on the electrodes. In order to model the experimental phenomena, herein, we simulate a simplified version of the electrodeposition problem on the electrode considering the necessary ingredients. In the electropolymerization process, one can expect monomers to become oligomers and higher sizes, which can be carrying the charge and hence driven by the electric field. Alternatively, one can consider the motion of concentration limited PSS$^-$ moving with the electric field, activating the growth of PEDOT on the electrodes. Irrespective of the identity of charge carrier (which is not clearly known), one can consider generic charge carriers represented as charged moving particles in the simulations. Further, the arguments of the model would be valid for systems in which nature of charge is opposite; in such a scenario, the deposition would occur on the opposite electrodes, maintaining similar morphologies. The advantage of the generic nature of the particles in the simulations makes the modeling applicable to electro-polymerized materials grown by different species. The electrical parameters applied on the electrodes can control the potential in the liquid. Further, the particles in the liquid phase can be expected to have several particle- particle and particle-fluid forces affecting the motion. Thus one needs to consider following ingredients in the model (1) potential distribution variation in the liquid due to time varying signal, (2) motion of particles governed by electrical parameters and various fluid interactions and (3) electro polymerization of the conducting polymer on the electrode. Figure 1c shows the proposed model with two bipolar metal electrodes shown in gray, and charge particles moving in the liquid shown in white. The electrodes are biased with the AC signal and spatiotemporal potential map is evaluated based on the Laplace equation using self-written codes in Fortran and Python (see Figure 1d). The particles' evolution is simulated based on a model that considers collective contributions of both field-effect electrodynamic drift and scattering (see Figure 1e). Consider a particle is at O and it would be experiencing unequal electrical fields in all directions and the random noise. Thus the particle would move in the direction where the collective drift based on both these effects (electrical-field assisted and thermally activated) would be higher. The probability of the motion is made proportional to the normalized value of drift along the direction. Since particles would be attracted towards the anode, and repelled by the cathode, thus their transience between both electrodes is ensured by the time varying nature of the voltage signal, according to the time constant of the monitored phenomenon and the frequency range of the applied signal. In the simulations, particles which happen to leave the boundaries are made to enter from the opposite direction at random positions to maintain constant concentration conditions ensured in the experimental setup. During the motion of particles with passage of time, the moving particles can happen to come near the electrode and once it comes near the electrode, it can polymerize and permanently attach to the electrode. Further, the potential value incorporates the drop in voltage at the double layer of the electrode with specific time constant as shown in Figure 1g, with red and blue color in contrast to the square signal shown with gray. The probability of particle-electrode attachment is made finite at a location where the particle electrode distance is < 2 pixels, enabling the contact from straight and diagonal direction. Without enabling the diagonal direction motion, the morphology is seen to be having rectangular artifacts. Further, since the conducting polymer polymerization is enabled by oxidation only if the anode reach the given oxidation potential, particle's potential shall be taken into account in the probability attachment to the anode. This is experimentally evidenced, with dendritic growth occurring above 3.5V as voltage amplitude, and the density of the morphology increases with the amplitude. To introduce voltage dependent oxidation, the attachment probability is made sigmoidal function, considering the voltage difference between the location of the particle and potential of double layer, higher the difference more is the attachment probability (representing oxidation). The stuck particles are made permanently attached to the electrode, as the polymerization process is irreversible. Since the polymerized particles on the electrode are conducting by nature, the deposited particles are considered integral part of the electrode with further oxidation (attachment) on dielectric surface occurring on the modified electrode surface (no voltage drop across the dendritic structure is taken into account in the model). The potential is recalculated based on the modified electrode geometry, and time varying signal on the electrode. The remaining particles in the liquid are simulated to move in the modified spatial-temporal potential field with the possibility of attachment on the modified electrode with the passage of time, enabling further growth of the polymer morphology. The polymer growth sensitively depends on the motion trajectory and attachment process, both regulated by the electrical conditions. Thus, versatility in the morphology can be seen based on electrical parameters. It should be noted that though in a real situation, the charge particles might be having negligible size as compared to the electrode spacing, and particle size may be variable, however, due to computational simplicity and cost restrictions, we assume the monomer or charge particle to have size of 1 pixel (1/20th of electrode spacing = 12 µm), and other length and time scale also based on simulation parameters. The present modeling parameters might be closer to the experimental system if one would be performing these experiments at much miniaturized conditions. Herein, we qualitatively, compare the model prediction with the experimental findings and predict the growth process in unexplored experimental conditions.

**Simulation parameters:**

| Experimental parameter (corresponding value) | Modeling parameter (corresponding value) |
|---|---|
| Electrode spacing (240 µm) | 20 pixels |
| Threshold voltage of electro polymerization ($V_A$) = 3.5 V | Threshold voltage of attachment ($V_A$) = 2 units |
| Charging time constant of double layer | Charging time ($t_c$) and Discharging time scale ($t_d$) = 5 simulation time units |
| Growth time scale (1s) | 100 Simulation time steps |
| Frequency of signal (Hz) | $(100 \times \text{Simulation time steps})^{-1}$ |
| 20 µL drop size | $150 \times 150$ pixels$^2$ |
| Charge particle size | 1 pixel |
| Concentration of moving charged particles (unknown) 10 mM of EDOT and (BQ) in 20 µL drop | number of particles (N) = 50 |
| Completion of growth: spacing between extreme points becoming zero | Spacing between extreme points is lesser than 8 pixels |
| Particle scattering due to fluid interaction | Random noise = 0.3 |
| Particle mobility (viscosity and particle dependent) | Mobility factor = 0.1 |

**Comparison of modeling results with experimental observations:**

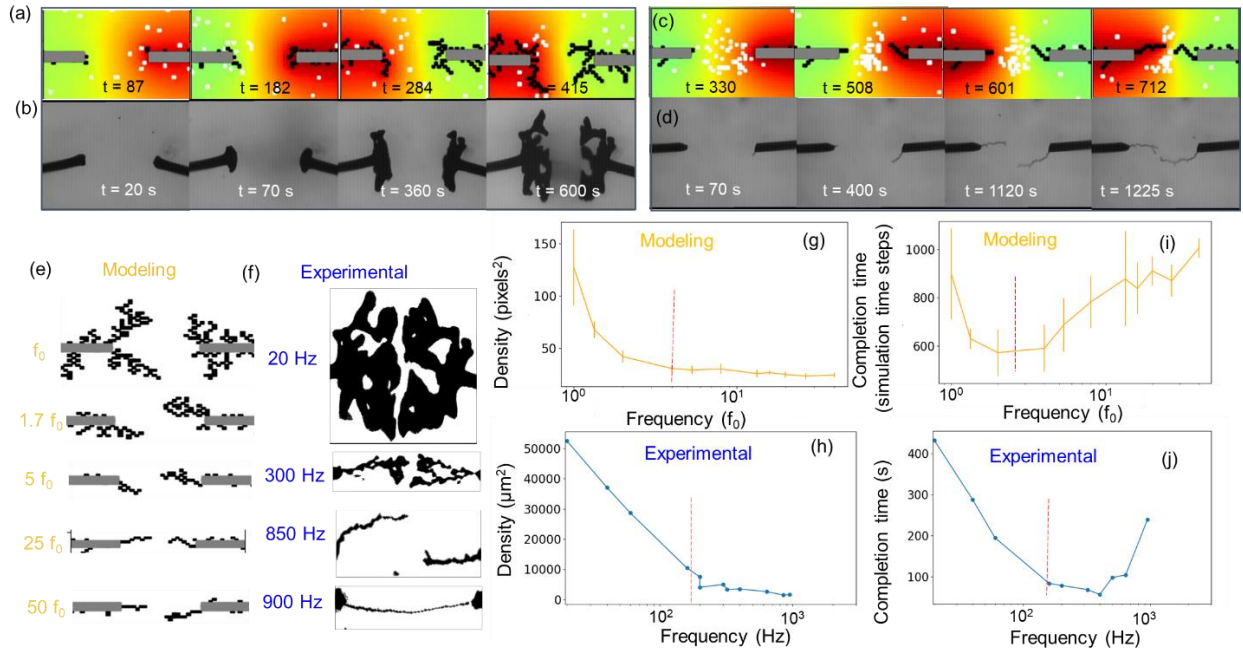

Figure 2: (a) Simulated images of dendrites growth process for signal frequency of 2.5 $f_o$ (low frequency). (b) Experimental time lapse images of dendrite structure with signal voltage of 5V and a low signal- frequency of 20 Hz. (c) Simulated images of dendrite morphology at signal frequency of 25 $f_o$ (high frequency) (d) Time lapse images of dendrites obtained from experiments at a high frequency value of 850 Hz. (e) The dendrite morphologies based on modeling for signal frequencies, $f_0$ to 50 $f_0$, here $f_0$ is defined in terms of simulation time steps of 100. (f) The microscopic binary images of dendrites morphologies obtained under variable frequency conditions ranging from 20 Hz to 900 Hz.[14] (g) Variation in the density values of modeled morphologies at different electrical signal frequencies. (h) Variation in 2D projected density for experimental morphologies. Variation in completion time for (i) modeled dendrites and (j) experimental dendrites.

Figure 2a shows the growth process observed from modeling prospective with a low frequency of 2.5 $f_o$ and 50 % duty cycle and zero offset. The morphology advances towards the opposite electrode with several branches making a fractal structure. Figure 2b represents a typical growth process obtained at experimental conditions of 5V, 20 Hz, 50 % duty cycle and zero voltage offset. The growth process takes 600 seconds, with initiation of growth in both the electrodes, followed diameter growth and fractal-like morphology growing towards the counter electrode. Furthermore, we see particle motions between the electrodes in the experimental video, which is consistent with the modeling assumption of particles moving between the electrodes. However keeping the conditions same, and only increasing the signal frequency one ends up having wire like growth for increased signal frequency (25 $f_o$). Growth starts from one end followed by growth from other electrode and moving towards each other. Figure 2d shows the experimental growth of dendrites at higher frequency (850 Hz) resembling the modeling growth process. Note that, the drastic change in fractal-to-wire-like growth cannot be trivially investigated in the experiments, given the distribution of moving particles in the liquid phase. In the modeling, (Figure 2c) we find that at increasing frequency, the particles tend to localize in the center of the electrodes vibrating minimally with low time period (high frequency) signal. The restricted particles' motion forces the growth only at the extreme tip of the dendrite, rather than in any other loation, offering wire like morphology with the passage of time. Figure 2e illustrates the images of dendrites obtained for a range of frequencies ($f_o$ to 50 $f_o$), the morphology is fractal and dense for $f_o$, the morphology becomes thin and less dense with increasing value of frequency, and becomes wire like at very high frequencies. The simulation well matches the experimental observations as shown in Figure 2f, as morphologies are fractal-like and very dense at 20 Hz, become less branchy and less dense with increasing frequency till 300 Hz, and become wire-like thereafter (see 800 Hz and 900 Hz). In Figure 2g, we plot the variation in the density of modeled dendrite morphologies obtained at different frequencies. One observes two regimes: (i) the decrease in density with increasing frequency from $f_o$ to $4f_o$, and (ii) saturation of density thereafter. In the experiments (Figure 2h) as well, it is observed that density drops systematically from 20 Hz to 300 Hz, while it saturates at a specific value from 300 Hz to 900 Hz. The completion time of the process is defined to quantify the kinetics of the growth and are considered when two dendrites are very close to each other. It is observed that the completion time in modeling and experiments (Figure 2i and 2j) both show similar nature; the completion time decreases with increase in frequency in regime I, while it increases in regime II. From the modeling we see that, In regime I, the increase in frequency helps in concentrating the particles near the center of electrodes. The high particle concentration increases the probability of dendrite growth and in turn decreases the completion time. However, in regime (ii) i.e. at very high frequency, the particles are dragged almost at the complete center of electrodes, vibrating minimally at low time-periods (high frequency). The restricted motion of these particles reduces the growth probability, and in turn slows down the process.

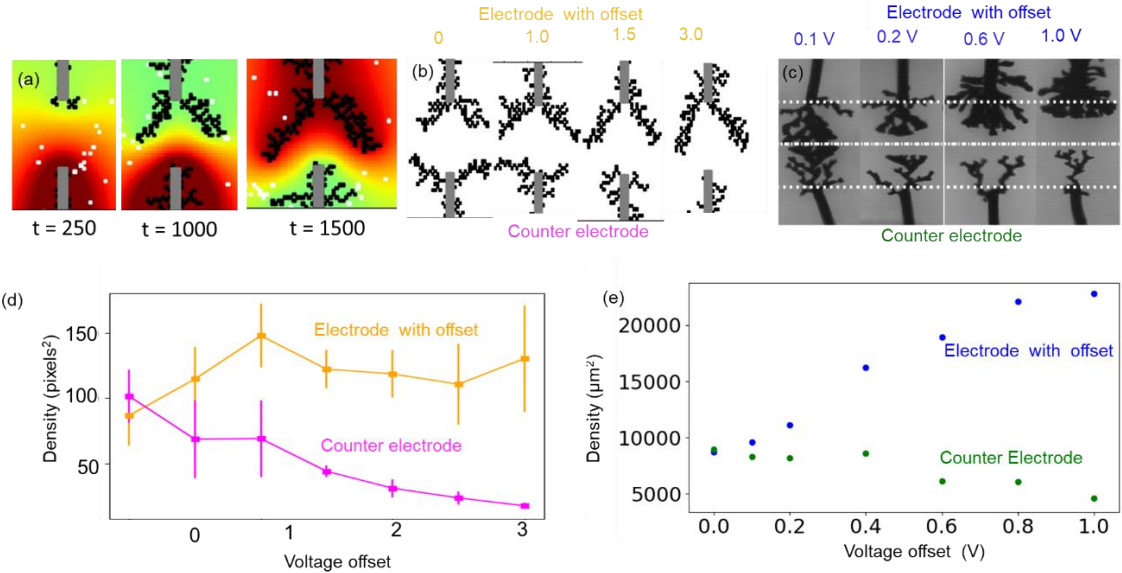

Figure 3: (a) The growth process of dendrites along with spatio-temporal map for voltage offset of 2 voltage units, demonstrating higher attraction of particles towards the electrode with offset. (b) Modeling images of morphologies for voltage offset of 0 to 3 voltage units. (c) Experimental images of dendrites morphologies at voltage offsets of 0.1 to 1 V.[14] (e) Dendrite density comparison for the modeling images for the electrode with offset (shown in orange) and counter electrode (shown in magenta). (e) Dendrite density comparison for the experimental images for the electrode with offset (shown in blue) and counter electrode (shown in green).

Next, we studied the effect of voltage offset, defined as a time invariant voltage component introduced in the applied periodic signal. Figure 3a shows the dendritic growth process with voltage offset of 2 units on the top electrode. An increased dendrite density is observed at the electrode that experience a positive voltage contribution by the voltage offset. The reason for this behavior can be explained based on increased density of particles and higher probability of attachment. Figure 3b-c compares the morphologies obtained from the modeling and the experimental methodologies at variable voltage offsets. The voltage offset increases asymmetry in both experiments and modeling. Figure 4d represents the systematic variation in the asymmetry with the increase in voltage offset for modeling images for electrode with offset (shown in orange) and counter electrode (shown in magenta). The similar trend is observed in the experimental studies[14] (Figure 3e), wherein the density on the electrode with offset (blue) increases while the density on counter electrode (green) decreases. In this way, the simplified model with few parameters can explain the morphologies obtained for different electrical signals.

## Modeling predictions:

Next, we studied parameters others than the voltage signal affecting the growth process of dendrites, which have not yet been systematically studied by experimentalists so far, and could be important points while designing the future experiments.

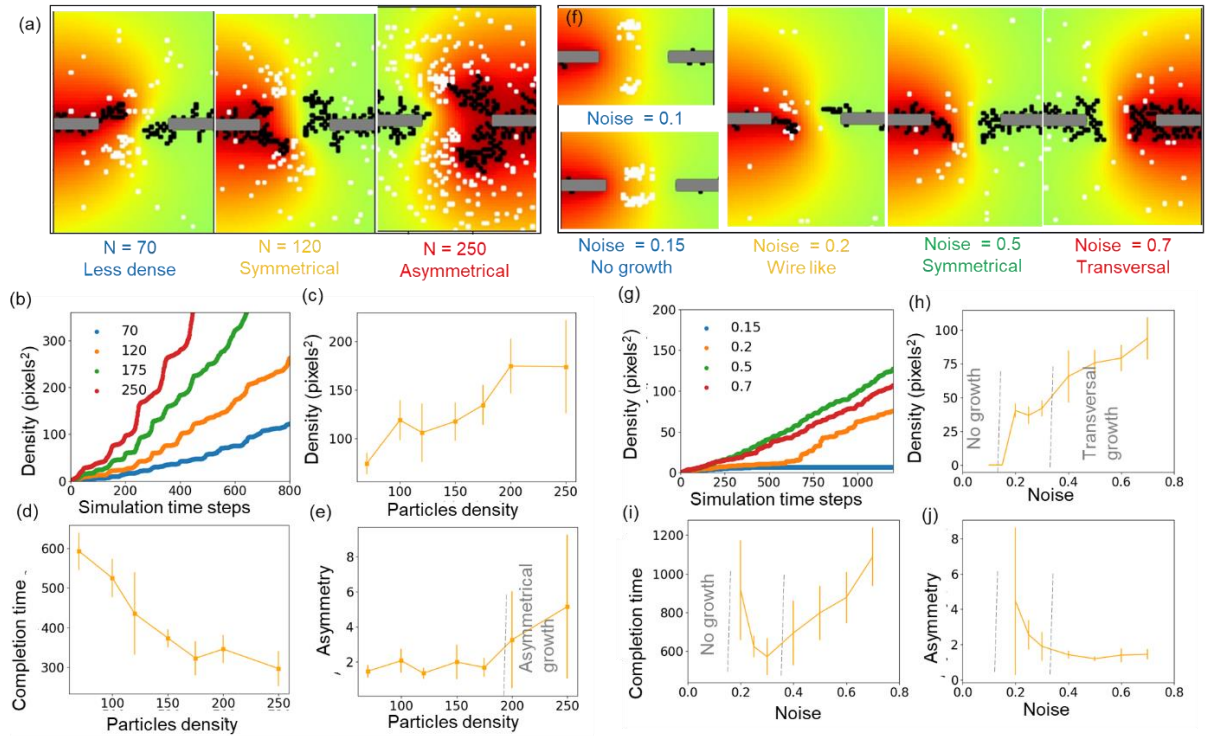

Figure 4: Effect of simulation parameters on the growth process. (a) The dendrites obtained at variable particle density N =70, 120 and 250. (b) The growth process for variable particle density. (c) The density of dendrites before being close to each other. (d) The variation in completion time with increase in particles density and (e) asymmetry parameter compared for dendrites obtained for different particle densities. (f) Various morphologies at increasing value of noise (η). (g) The time evolution of dendrites' density for variable noise values. (h) The density of dendrites before completion and (i) corresponding completion time for variable noise values. (j) Asymmetry parameter compared for dendrites obtained at different noise values.

The monomer concentration is an important aspect in experiments that affects dendritic growth. Figure 4a shows the dendritic growth at increasing particle densities, demonstrating very distinct morphologies for different particle densities: (i) wire-like for N = 70, (ii) fractal and symmetrical at N = 120, and (iii) asymmetrical at N = 250. The findings are consistent with the experimental observations of Ohira et al. wherein, wire like dendrites were obtained in the reduced concentration by glass covering the region during bipolarization process.[34] Based on the modeling it can be explained based on more attachment events at high particle population. However, at very high particle density, the structure becomes asymmetrical, despite having similar conditions on both electrodes. Because even a small noise starts the growth in any one of the electrodes, which acts as the nucleating center and triggers attachments of neighboring particles for further growth. Since at higher density, in a small-time interval, many particles can attach to the nucleating center, making further nucleating centers and growth of dense structure. Thus, high particle density can induce asymmetrical growth and noise-induced winner-takes-all kind of situation. The time evolution of the growth process shown in Figure 4b strengthens the claim as the growth is uniform for lower particle densities, while the growth process becomes fast and discontinues and step like for higher particle densities, manifesting growth of several particles in a small interval. An increase in particle density, increases the possibilities of attaching and in turn increases the density as shown in Figure 4c, rises from N = 70 to N = 200 and saturates further. The growth rate can be quantified in terms of completion time as shown in Figure 4d, drops systematically from N 70 to N= 175, and nearly becomes constant further. A term asymmetry is defined in Figure 4e to compare dendrites' morphologies originating from both the electrodes, with the value corresponding to the ratio of dendrite with high density to lower density. The minimum possible value of 1 corresponds to complete symmetrical structure. It is observed that the value is below 2 for dendrites N = 70 to N = 150, while the value rises significantly further with an asymmetry value above 4 for N = 250. Thus, without any asymmetry in electrical parameters, one can end up in asymmetrical structures based on intrinsic growth processes with higher particle densities.

The motion of the particles are regulated by electric field and thermally activated random motion (scattering). To understand the impact of scattering as a random contribution in the denditic growth, the growth has been studied at different values of noise ($\eta$) at other constant signal conditions. If we see the effect of noise on the morphologies (Figure 4f), at very low noise ($\eta$ <0.2), there is no growth on any of the electrodes, with the increasing noise the growth turns to wire like ($\eta$ = 0.2), and with medium noise the growth is bulk fractal structure ($\eta$ = 0.5). On the other hand, for very high noise ($\eta$ = 0.7), the growth starts occurring in the electrode wire in the transversal direction. The reasons for these effects can be explained based on difference in particle distribution at variable noise levels. Due to relatively higher value of electric field at low noise levels, the particles are solely governed by the field, since the signal changes the polarity based on periodic signal, hence the particles concentrate near the center. At very low noise, the particles are frozen in the mid of electrodes ($\eta$ <0.2), leading to no growth. With increasing contribution of the noise, particles can move from the center towards the electrode. At $\eta$ = 0.2, only a few particles can reach the electrode, it is similar to the case of less concentration (say N = 70) studied previously, and hence wire-like growth is manifested. The increase in noise ($\eta$ = 0.5) increases the particle density and hence fractal structure is obtained. However, at very high noise, the motion is not controlled by the field, and high density of particles get available throughout the system, increasing the transversal growth. The growth process of wire ($\eta$ = 0.2), plotted in Figure 4g shows that the wire growth takes certain threshold time to begin with. The dendrite density and completion time plots (Figure 4h-i) also shows distinct regimes, no growth for ($\eta$ <0.2), high variability in the growth time and asymmetry for $\eta$ = 0.2, reduction in completion time for medium noise ($\eta$ 0.2- 0.3) since the particles can easily approach the electrodes, for high Noise (0.3-0.7), the growth on radial of wire starts manifesting, and completion time increases.

Thus, as previously noted, the influence of noise, which is dependent on mass diffusibility, can result in a variety of states. Because diffusibility is determined by the mass and particle-solvent interaction, various monomers might result in varied morphological morphologies, which is an essential consideration for experiments throughout the optimization process. At high concentration, particle concentration can also play a role in converting wire-like growth to bulk fractal-like growth and asymmetrical growth, which should be considered while experimenting with concentration in the optimization process. Apart from neuromorphic engineering community, the investigation can be an important contribution to various other domains wherein such interconnections and morphological shape control their functionalities. [22-29] The modeling can also be explored for pattering applications[35] and locomotion of conducting objects with similar techniques[36-37]. The modeling can also be translated to problems involving wireless electro-polymerization.[16, 38] Since we used generic charged particles in the modeling, the model and its findings would also be important to inorganic electrochemical depositions. [33-34,39] Future work can be done with non-equal mesh size for potential determination to have fine resolution in potential map near the dendrites. Currently, due to different time scales and length scale of simulations as compared to experiments, we are not able to provide one to one mapping. Future work could attempt in relating the modeling parameters into the experimental units. The modeling can also be extended for network-based simulations by introducing multiple electrodes in the modeling. Currently, we have shown the simulations for periodic signals, the approach can also be applied on non-periodic time series for neuromorphic applications. We believe that the present modeling foundations and inferences from the modeling studies would be helpful in understanding and optimizing the electrode geometries and electrical signals for the upcoming neuromorphic devices.

## Conclusion

In conclusion, we have developed a mesoscale model to understand and optimize the electro polymerization technique as a bottom up strategy for neuromorphic computation and other applications. The model involves consideration of spatial-temporal potential mapping for time varying signal across the electrodes, motion of charged particles in the electric field, attachment of particles to the electrode. The model attempts to explain the morphological differences for different electrical activities and experimental conditions, based on the nature of particles distribution and their trajectories for electrical activities, not studied before via modeling or microscopic technique.  The increasing frequency turns a bulky structure into a wire like morphology due to preferential distribution of particles near the center of electrodes and tip of the growth. The voltage offset brings the asymmetry in the morphology due to asymmetry in the particle distribution due to constant potential. The effect of concentration shows different regions, which are wire like, bulky and asymmetrical at increasing concentration values due to the increases time scale of the growth process. The effect of scattering can also a have a huge impact, with no growth, wire like growth, symmetrical growth and transversal growth at increasing value of scattering due to reduction of relative electric field contribution. The model can be explored for multiple electrodes and making quantitative experimental comparisons after further modifications.

## Acknowledgements:

The authors wish to thank the European Commission: H2020-EU.1.1.ERC project IONOS (# GA 773228).